\documentclass[sigconf]{acmart}

\AtBeginDocument{%
  }

\setcopyright{acmlicensed}
\copyrightyear{2026}
\acmYear{2026}
\acmDOI{XXXXXXX.XXXXXXX}
\acmConference[EASE 2026]{The 30th International Conference on Evaluation and Assessment in Software Engineering}{9–12 June, 2026}{Glasgow, Scotland, United Kingdom}

\usepackage{tabularx}

\begin{document}

\title{TDD Governance for Multi-Agent Code Generation via Prompt Engineering}

\author{Tarlan Hasanli}
\authornote{Both authors contributed equally to this research.}
\email{tarlan.h.hasanli@jyu.fi}
\orcid{0009-0002-4578-5908}
\affiliation{%
  \institution{University of Jyväskylä}
  \city{Jyväskylä}
  \country{Finland}
}
\author{Shahbaz Siddeeq}
\authornotemark[1]
\email{shahbaz.siddeeq@tuni.fi}
\orcid{0009-0003-9030-8841}
\affiliation{%
  \institution{Tampere University}
  \city{Tampere}
  \country{Finland}
}

\author{Bishwash Khanal}
\orcid{0009-0008-6972-4660}
\email{bishwash.b.khanal@jyu.fi}
\affiliation{%
  \institution{University of Jyväskylä}
  \city{Jyväskylä}
  \country{Finland}
}

\author{Pyry Kotilainen}
\orcid{0000-0002-4645-074X}
\email{pyry.kotilainen@jyu.fi}
\affiliation{%
  \institution{University of Jyväskylä}
  \city{Jyväskylä}
  \country{Finland}
}

\author{Tommi Mikkonen}
\orcid{0000-0002-8540-9918}
\email{tommi.j.mikkonen@jyu.fi}
\affiliation{%
  \institution{University of Jyväskylä}
  \city{Jyväskylä}
  \country{Finland}
}

\author{Pekka Abrahamsson}
\email{pekka.abrahamsson@tuni.fi}
\orcid{0000-0002-4360-2226}
\affiliation{%
  \institution{Tampere University}
  \city{Tampere}
  \country{Finland}
}

\renewcommand{\shortauthors}{Hasanli et al.}

\begin{abstract}
Large language models (LLMs) accelerate software development but often exhibit instability, non-determinism, and weak adherence to development discipline in unconstrained workflows. While test-driven development (TDD) provides a structured Red-Green-Refactor process, existing LLM-based approaches typically use tests as auxiliary inputs rather than enforceable process constraints. We present an AI-native TDD framework that operationalizes classical TDD principles as structured prompt-level and workflow-level governance mechanisms. Extracted principles are formalized in a machine-readable manifesto and distributed across planning, generation, repair, and validation stages within a layered architecture that separates model proposal from deterministic engine authority. The system enforces phase ordering, bounded repair loops, validation gates, and atomic mutation control to improve stability and reproducibility.
We describe architecture and discuss encoding software engineering discipline directly into prompt orchestration, which we think offers a promising direction for reliable LLM-assisted development.
\end{abstract}



\maketitle

\section{Introduction}

The discipline of software engineering is undergoing a major shift as autonomous LLM-based agents take on complex coding tasks \cite{vibecodingvsagentic}. In agentic AI, role-specific agents collaborate to achieve programming goals with limited human intervention \cite{surveycodegenerationllmbased}. While this can improve productivity, LLMs remain non-deterministic. Even identical prompts often produce different outputs, and a temperature of zero does not ensure consistent results \cite{10.1145/3697010}. In multi-agent settings, small logic errors can spread across the workflow, which makes automated guardrails essential for code quality and reliability \cite{surveyvibecodinglarge, agenticreasoninglargelanguage}.

One potential solution comes from Test-Driven Development, introduced through Extreme Programming, where tests are written before implementation so correctness is defined in advance \cite{beck2002}. However, TDD can be demanding because developers must maintain both design intent and detailed test logic, which often reduces productivity during adoption \cite{5770623, staegemann2022}. AI-native engineering offers a practical complement: TDD provides clear pass or fail constraints, and LLMs can generate much of the repetitive test code that developers often find tedious \cite{fakhoury2024exploring}.

This paper presents a prompt-centric multi-agent framework that turns TDD principles into enforceable rules for LLM-based code generation. Specialized agents follow explicit guidance, and an orchestrator controls phase order and checks test outcomes before transitions. It also limits repair loops. By embedding TDD into the prompt architecture, the framework aims to reduce instability in LLM-generated code while preserving the efficiency benefits of agentic AI

\section{Background and Motivation}


This section reviews the empirical evidence that motivates encoding TDD as a governed multi-agent protocol. We organize the discussion around three themes: classical TDD foundations, LLM instability factors, and test-guided prompting approaches.

Beck's formulation of TDD operationalizes development as micro-iterations: tests specify behavior before implementation, encourage minimal solutions, and require continuous refactoring after reaching a passing state \cite{beck2002}. Fowler's refactoring catalog complements this by defining safe structural transformations and highlighting that automated tests are the prerequisite for verifying behavior preservation \cite{fowler2018refactoring}. Together, these sources motivate a process view in which tests are not merely evaluation artifacts but an explicit control mechanism for incremental design change.

Recent advances in large language models have reduced the cost of code generation, enabling  synthesis from natural-language specifications and interactive debugging. Test-guided prompting improves functional correctness and developer confidence in LLM-assisted tasks \cite{10.1145/3691620.3695527, fakhoury2024exploring}, but these systems do not inherently enforce disciplined processes such as phase ordering, minimal implementation, or bounded repair cycles. Repository-scale analysis shows that LLMs produce incorrect or nonsensical code due to contextual dependency errors, insufficient grounding, and overconfident interpolation \cite{zhang2025llm}. Retrieval augmentation reduces hallucinations, indicating that reliability depends on prompt context quality as well as model capability.

Non-determinism further weakens workflows stability: identical prompts often produce different outputs, with zero equal test outputs in up to 75.76\% of cases on complex benchmarks \cite{10.1145/3697010}. Similar variability appears in code review tasks \cite{klishevich2025measuring},  and apparently correct code may still fail in clean environments due to dependency and configuration issues \cite{vangala2026aigeneratedcodereproducibleyet}. These findings suggest that functional correctness alone is insufficient and motivate governed workflows with reproducibility controls.

Interaction-centered evidence shows further failure modes, including incomplete answers, excessive outputs, missing preconditions, and prompt sensitivity, often requiring manual decomposition and iterative steering \cite{tie2024llmsimperfectwhatempirical}. Test cases improve performance by reducing ambiguity and acting as executable specifications \cite{10.1145/3691620.3695527}. WebApp1K similarly finds that when tests serve as both a prompt and a verifier, success depends heavily on instruction following and in-context learning \cite{cui2025tests}.

Interactive workflows further support test-centric iteration. TiCoder blends TDD-style test refinement with code generation, reporting improved correctness and reduced cognitive load \cite{fakhoury2024exploring}. 
At repository scale, TDFlow decomposes workflows into specialized sub-agents for patching, debugging, and revising, achieving high pass rates with ground-truth tests \cite{han2025tdflow}. Our contribution instead emphasizes governed execution under model non-determinism through deterministic validation, bounded repair, and engine-controlled state mutation.
Practical guidance for TDD-style LLM use emphasizes that test-first interaction patterns, descriptive signatures, and focused unit tests improve reliability over naive prompting \cite{10.1145/3643795.3648382}.

Empirical prompt engineering work shows that iterative self-evaluation loops (Self-Refine) and structured Chain-of-Thought prompting measurably improve code generation quality \cite{shin2025promptengineeringfinetuningempirical, li2025structured}. Integrating deterministic diagnostics, such as static analysis warnings as feedback, can further improve non-functional dimensions like maintainability \cite{santana2025promptingtechniqueiuse}.

These studies establish that structured prompting and deterministic feedback improve outcomes, but they do not specify how to enforce process discipline, specifically TDD phase ordering, minimal implementation, and bounded repair, across multi-agent workflows. Three empirically supported observations summarize this gap:
\begin{enumerate}
    \item Classical TDD relies on disciplined phase ordering and behavior-preserving refactoring backed by tests \cite{beck2002, fowler2018refactoring}.
    \item LLM-based software engineering exhibits instability through hallucinations, non-determinism, and reproducibility gaps \cite{zhang2025llm, 10.1145/3697010, klishevich2025measuring, vangala2026aigeneratedcodereproducibleyet}.
    \item Test-guided and structured prompting approaches improve correctness but do not enforce full TDD discipline as a governed protocol \cite{10.1145/3691620.3695527, cui2025tests, fakhoury2024exploring, han2025tdflow, 10.1145/3643795.3648382}.
\end{enumerate}

A central open problem therefore remains: how to translate classical TDD principles into enforceable, governance-aware prompt structures that ensure bounded autonomy and phase discipline across multi-agent workflows. The framework presented in Section~\ref{sec:design} addresses this gap.
\section{AI-Native TDD} 
\label{sec:design}


\subsection{Deriving Enforceable Principles from TDD Literature}


We begin by describing how TDD principles are extracted and formalized as enforceable guidelines. Our goal is not to summarize TDD books or re-document practices. Instead, we distill the key normative rules that early TDD literature treats as correct. We focus on rules that often break down in real teams under time and delivery pressure. We therefore focus on principles that were framed as the "right" way to develop, yet were economically fragile for humans because they demanded continuous attention, constant feedback, and non-deferrable cleanup work.


\textit{Source scope and selection.} We bounded the extraction corpus to Kent Beck~\cite{beck2002} and Robert C. Martin~\cite{martin2009clean, martin2011clean} as a small canonical and practice-oriented source base. These texts were selected because they articulate TDD in explicit prescriptive terms, center the Red-Green-Refactor micro-cycle, and frame rapid, trustworthy feedback as the core mechanism for reducing development risk. The objective of this step was not to synthesize all influential TDD perspectives, but to derive a methodologically consistent set of norms that could be translated into enforceable governance rules for AI-assisted development. Other influential perspectives, including later design-oriented or object-oriented interpretations of TDD, were therefore not excluded because they lack relevance, but because incorporating them would have required a broader comparative synthesis across partially heterogeneous formulations of TDD, which was outside the scope of this study. We accordingly treat the resulting principle set as corpus-bounded rather than exhaustive. We treated statements as candidates for extraction when they satisfied two criteria: (i) the statement is prescriptive (a “must/should” about how to work), rather than descriptive or tool-specific; and (ii) violating the statement is a historically common coping strategy under deadline pressure (e.g., batching changes, weakening tests, or skipping refactoring).

\textit{Extraction method.}
We performed a manual, concept-level synthesis rather than a chapter-by-chapter summary. We collected candidate statements and normalized them into a consistent principle form with: (1) a short label, (2) an "original intent" phrased in human-era terms (why this was believed right), and (3) a source reference. We merged them into a single principle in which multiple phrasings expressed the same underlying norm. Conversely, we split it into separate principles to preserve machine-actionable semantics when a single passage contained multiple independent norms (e.g., phase ordering plus minimality).

\textit{What qualifies as a "principle."}
We define a TDD principle as a constraint on \emph{order}, \emph{granularity}, or \emph{quality of feedback} in the development loop: \textbf{Order constraints} enforce sequencing (e.g., test-first, Red--Green--Refactor), \textbf{Granularity constraints} enforce small steps (e.g., one failing test at a time, minimal passing code), and \textbf{Feedback-quality constraints} ensure feedback remains fast and trustworthy (e.g., tests must be fast and deterministic; assertions must be meaningful; refactoring must preserve behavior). Table~\ref{tab:principle_categories} lists these categories and the typical human-era failure mode that causes each category to degrade under delivery pressure.
(3) 




\begin{table}[!htb]
\centering
\footnotesize 
\setlength{\tabcolsep}{3.5pt} 
\renewcommand{\arraystretch}{1.1}

\begin{tabularx}{\columnwidth}{>{\bfseries}p{1.9cm} X X}
\hline
Category & Canonical principle form & Why humans dropped it under pressure \cite{5770623, staegemann2022} \\
\hline
Order & Test-first and Red$\rightarrow$Green$\rightarrow$Refactor 
& Front-loads thinking and "just code it" feels faster \\

Granularity & Minimal failing test, minimal passing code, and one failing test at a time 
& Batching reduces perceived overhead but increases hidden risk \\

Feedback quality & FAST, independent, repeatable, self-validating, and timely tests with meaningful assertions 
& Slow or flaky suites and weak tests become socially tolerated to ship \\

Design hygiene & Remove duplication and refactor continuously while green 
& Benefits are delayed and invisible, so refactor becomes optional \\
\hline
\end{tabularx}

\caption{Principle categories and their typical human-era failure modes.}
\label{tab:principle_categories}
\end{table}

\textit{Historically "right" but economically fragile.}
The extracted set intentionally over-represents principles that are difficult for humans to sustain. For example, "run tests constantly" and "refactor as a required step" are widely accepted, but frequently abandoned because the short-term cost is salient while the benefit is delayed. Similarly, "tests must be self-validating" is accepted in principle, yet humans commonly ship weak tests (or non-assertive tests) because they are faster to write. In other words, the extraction targets the \emph{known failure points} of human-era TDD: principles that require repeated, immediate investment to preserve long-term optionality.

\begin{figure}[!htbp]
    \centering
    \includegraphics[width=\columnwidth, clip, trim=0 20 0 20]{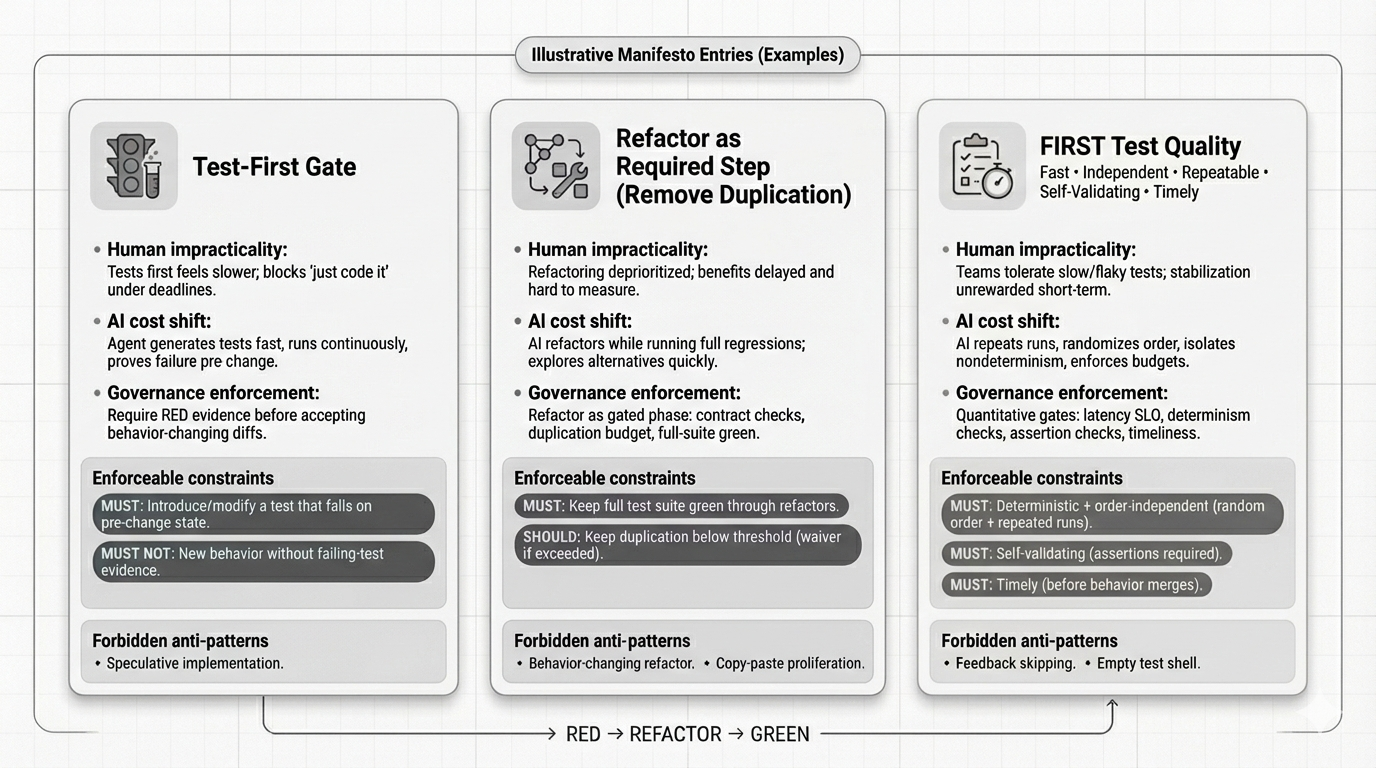}
    \caption{Illustrative manifesto entries and their governance structure.}
    \label{fig:Manifesto}
\end{figure}

\textit{Outputs and representation.}
The outcome of this step is a curated list of principles representing the bounded canonical TDD corpus used in this study. Each principle is represented as a structured record containing: a label, a canonical quote (when available), a human-era intent statement, and a bibliographic pointer. This representation is designed as the input to the next step: translating human-era discipline into AI-native governance rules that autonomous or semi-autonomous systems can ingest and enforce. Figure \ref{fig:Manifesto} illustrates three enforceable governance principles: requiring pre-change failing tests (RED) for behavior modifications, treating refactoring as a gated phase with full-suite regression and duplication controls, and operationalizing FIRST test quality via quantitative checks for speed, determinism, self-validation, and timeliness.

\textit{Manifesto representation.} The machine-readable manifesto is implemented as a JSON-based structured\footnote{\url{https://github.com/shahbazsiddeeq/TDD-manifesto/blob/main/tdd_principles_manifesto.json}} schema in which each principle is encoded as a governance object. In addition to bibliographic grounding, each entry includes an identifier, a concise title, the original human-oriented intent, an AI-native interpretation, operational constraints, and anti-patterns. This representation is used at two levels: interpretation fields guide role-specific prompt construction, while constraint and anti-pattern fields are mapped to runtime enforcement points such as phase gating, scope restriction, proposal validation, and bounded repair control. Figure \ref{fig:Manifesto} illustrates representative manifesto entries and their translation into enforceable governance elements.

\subsection{Proposed Approach and System Description}


We operationalize classical TDD as a governed, multi-phase orchestration framework in which TDD principles are encoded as structured constraints distributed across planner, generation, repair, and validation stages. Rather than relying on ad hoc prompting or post hoc verification, the system enforces phase ordering, bounded autonomy, and deterministic validation at each step of the development loop. Large language models act as proposal generators, while an execution engine maintains authority over state mutation and workflow progression. Unlike TDFlow-style autonomous execution, the orchestration engine retains sole authority over workspace mutation.

\textit{Phase-Oriented workflow.} The system follows a structured execution cycle aligned with the Red-Green-Refactor discipline. A specification is first decomposed into ordered steps by a planner component. These steps explicitly encode expected test outcomes, ensuring that failing tests (red) lead to implementation (Green). Code generation is therefore not permitted until a failing test state is established.

\begin{figure}[htbp]
    \centering
    \includegraphics[width=\columnwidth, clip, trim=3cm 19cm 8cm 1.3cm]{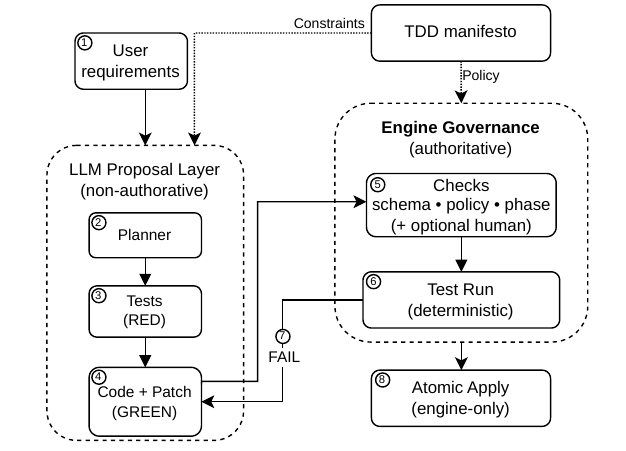}
    \caption{Governed AI-native TDD workflow with separation between non-authoritative proposal generation and authoritative execution.}
    \label{fig:architecture}
\end{figure}

After each proposal, the system executes a review and validation stage before applying any changes to the project workspace. Only after successful validation are changes applied and tests executed. If tests fail, the system enters a repair loop; if tests pass, the workflow proceeds toward completion and optional refactoring (Figure \ref{fig:architecture}).


\textit{Governance-Centric Architecture.} Beyond sequential phase ordering, the system is designed around a governance-centric architecture that separates generative autonomy from state authority. Language models never directly write to the file system. Instead, they return structured patch proposals that are subject to validation gates prior to application.

The system incorporates a structured TDD manifesto that encodes extracted principles as declarative constraints. These constraints inform prompt construction, validation rules, and phase enforcement logic, serving as the conceptual contract between TDD theory and runtime orchestration.

Before any mutations occur, outputs undergo:

\begin{enumerate}
    \item Structural validation (schema and content checks),
    \item Policy enforcement (e.g., disallowed paths, directory restrictions),
    \item Phase consistency checks (ensuring outputs match the expected phase),
    \item Optional human or rule-based approval (in planner mode). 
\end{enumerate}

Only after passing these gates are changes applied atomically. All file mutations are performed exclusively by the orchestration engine after validation, not by the model. This separation ensures that generative variability does not directly translate into uncontrolled state changes (Figure \ref{fig:architecture}).

Figure~\ref{fig:architecture} illustrates the end-to-end governed workflow. First, user requirements and TDD manifesto constraints are provided as inputs. The proposal layer then generates an execution plan through the planner. During the RED phase, failing tests are proposed to establish the expected behavior. In the GREEN phase, implementation proposals (code patches) are generated. All proposals are routed through the engine governance layer, where schema validation, policy enforcement, and phase-consistency checks are applied. Approved proposals are executed through a deterministic test run. If tests fail, control returns to the proposal layer for bounded repair iterations; otherwise, validated changes are atomically committed to the workspace.

\textit{Prompt Distribution Across Agent Roles.} TDD discipline is not encoded in a single instruction. Instead, it emerges from coordinated constraints distributed across role-specific prompts. \textbf{System prompt} establishes governance invariants, structured output requirements, and phase purity. \textbf{Planner prompt} decomposes the specification into ordered steps with expected test outcomes (e.g., FAIL then PASS), thereby encoding test-first progression and constraining phase transitions. \textbf{Test Generation prompt} enforces RED-phase constraints, restricting output to the test files and requiring meaningful assertions aligned with the specification. \textbf{Implementation prompt} restricts synthesis to minimal changes necessary to satisfy failing tests and forbids unrelated feature additions. \textbf{Failure Repair prompt} injects structured failure context (e.g., failure category and test output) and enforces minimal, localized corrections. \textbf{Review prompt} acts as a quality gate, prohibiting production edits during test review and preventing over-specification or invented requirements.

Through this distribution, classical TDD properties - test-first execution, minimal implementation, and behavior preservation - are enforced at both prompt and engine levels.

\textit{Bounded Repair and Loop Control.}



Repair is governed by explicit termination policies to ensure convergence. Each GREEN step allows at most $N=3$ repair attempts, chosen as a fixed retry budget to bound cost while preserving iterative recovery. A failure signature $S = \{\text{exception type}, \text{failing tests},\\ \text{normalized message}\}$ is computed per iteration. The loop terminates early if: 
\begin{enumerate} 
    \item the same $S$ repeats in consecutive iterations,
    \item a repair produces no effective code change (detected via patch comparison), or 
    \item proposals are semantically equivalent to prior attempts. Execution also terminates when tests pass or the iteration cap is reached.
\end{enumerate}
Execution also terminates when tests pass or the iteration cap is reached. These constraints ensure bounded, reproducible repair behavior.

\textit{From Principles to Enforceable Mechanisms.}
The extracted TDD principles are translated into concrete enforcement mechanisms within the orchestration pipeline:

\begin{itemize}
    \item \textbf{Test-first development} is enforced through planner-ordered steps and engine-level FAIL gating, preventing code generation before failing tests exist.
    \item \textbf{Minimal implementation} is reinforced by prompt-level scope restrictions and engine rejection of no-op or unrelated changes.
    \item \textbf{Behavior preservation} is guaranteed by post-apply test execution gates and automatic rollback during refactoring if tests fail.
    \item \textbf{Controlled refactoring} is permitted only after successful test passes and is subject to validation that forbids feature modification or test edits.
\end{itemize}

Importantly, these guarantees arise from the interaction between prompt constraints and deterministic validation rules. TDD is therefore not advisory but operationalized as a distributed governance protocol.

\textit{AI-Native TDD as Process Encoding}

Unlike approaches that treat tests as auxiliary inputs or evaluation metrics, this framework encodes TDD as a process-level constraint architecture. Phase ordering, validation gating, bounded repair, and state authority are integrated into the generative loop itself. As a result, the development discipline shapes the trajectory of model generation rather than being applied after the fact.

This design positions prompt engineering not merely as task phrasing but as a mechanism for encoding software engineering process invariants within AI-assisted development workflows.

\section{Discussion and Future Work}


The primary benefit of the AI-native TDD framework is the explicit encoding of development discipline within the generative workflow. By separating proposal generation from state mutation and enforcing phase-ordered execution, the system reduces uncontrolled iteration and mitigates instability common in LLM-driven development. TDD principles are distributed across planner, generation, repair, and validation stages, enabling process-level governance beyond prompt phrasing alone. Bounded repair loops and deterministic validation gates improve reproducibility, while the structured TDD manifesto operationalizes test-first development as a runtime constraint, preserving behavioral safety during iterative synthesis.


The current implementation has several limitations. Manifesto-derived constraints are primarily enforced at the prompt level, with only partial runtime verification, leaving full semantic compliance as future work. While the bounded autonomy model stabilizes execution, it may limit exploration in complex refactoring scenarios. Empirical validation remains preliminary, and broader cross-model and repository-scale evaluations are needed to assess the robustness. Additionally, scaling to large, multi-module repositories with complex dependencies will require more advanced planning and invariant enforcement mechanisms.


Initial experimentation suggests that explicit phase separation and validation gating reduce unstable retry cycles compared to baseline prompting. Enforcing test-first ordering and minimal implementation appears to limit speculative code generation and unnecessary feature expansion. However, manifesto injection must be carefully balanced against token constraints, highlighting a trade-off between governance strength and prompt compactness.


Future work will evaluate AI-native TDD in repository-scale industrial settings, including CI/CD integration, to assess stability, defect rates, and reproducibility. We also plan to explore configurable governance levels that allow teams to calibrate strictness based on project complexity. The deterministic validation and bounded autonomy model may further support auditable AI-assisted development, particularly in regulated domains.



\begin{acks}
This work has been supported by Business Finland (projects GENIUS (2545/31/2024) and ANSE (1822/31/2025)). We would also like to acknowledge the use of Google’s Nano Banana Pro in generating the visualization for Figure \ref{fig:Manifesto}.
\end{acks}

\bibliographystyle{ACM-Reference-Format}
\bibliography{references.bib}

\end{document}